\documentclass[vecphys]{svmult}

\usepackage{makeidx}         
\usepackage{graphicx}        
\usepackage{multicol}        
\usepackage[bottom]{footmisc}
\usepackage{amsfonts}
\usepackage{amsmath}
\usepackage{amssymb}
\usepackage{epsfig}
\usepackage{graphicx}
\usepackage{latexsym}
\usepackage[pdftex]{hyperref}
\usepackage{xy}
\usepackage{pstricks}
\usepackage{bm}

\makeindex             

\begin{document}

\title*{Non-Supersymmetric Attractors in Symmetric Coset Spaces}
\author{Wei Li}
\institute{Jefferson Physical Laboratory, Harvard
University, Cambridge MA 02138, USA
\texttt{weili@fas.harvard.edu}
}
%
%
\maketitle

\section{Introduction}

The attractor mechanism for supersymmetric (BPS) black holes was discovered in 1995 \cite{Ferrara:1995ih}: at the horizon of a supersymmetric black hole, the moduli are completely determined by the charges of the black hole, independent of their asymptotic values. In 2005, Sen showed that all extremal black holes, both supersymmetric and non-supersymmetric (non-BPS), exhibit attractor behavior \cite{Sen:2005wa}: it is a result of the near-horizon geometry of extremal black holes, rather than supersymmetry. Since then, non-BPS attractors have been a very active field of research (see for instance \cite{Goldstein:2005hq,Sen:2005iz,Tripathy:2005qp,Alishahiha:2006ke,Kallosh:2006bt,Chandrasekhar:2006kx,
Sahoo:2006rp,Astefanesei:2006dd,Kallosh:2006ib,Sahoo:2006pm,Andrianopoli:2006ub,D'Auria:2007ev,Nampuri:2007gv}).  In particular, a microstate counting for certain non-BPS black holes was proposed in \cite{Dabholkar:2006tb}. Moreover, a new extension of topological string theory was suggested to generalize the Ooguri-Strominger-Vafa (OSV) formula so that it also applies to non-supersymmetric black holes \cite{Saraikin:2007jc}.

Both BPS and non-BPS attractor points are simply determined as the critical points of the black hole potential $V_{BH}$ \cite{Ferrara:1997tw,Kallosh:2006bt}. However, it is much easier to solve the full BPS attractor flow equations than to solve the non-BPS ones: the supersymmetry condition reduces the second-order equations of motion to first-order ones. Once the BPS attractor moduli are known in terms of D-brane charges, the full BPS attractor flow can be generated via a harmonic function procedure, i.e., by replacing the charges in the attractor moduli with corresponding harmonic functions:
\begin{equation}\label{kharmonic}
t_{BPS}(\vec{x})=t_{BPS}^{*}(p^I\rightarrow H^I(\vec{x}),q_I\rightarrow H_I(\vec{x}))
\end{equation}
In particular, when the harmonic functions $(H^I(\vec{x}),H_I(\vec{x}))$ are multi-centered, this procedure generates multi-centered BPS solutions \cite{Bates:2003vx}.

The existence of multi-centered BPS bound states is crucial in understanding the microscopic entropy counting of BPS black holes and the exact formulation of OSV formula \cite{Denef:2007vg}. One can imagine that a similarly important role could be played by multi-centered non-BPS solutions in understanding non-BPS black holes microscopically. However, the multi-centered non-BPS attractor solutions have not been constructed until \cite{Gaiotto:2007ag}, on which this talk is based. In fact, even their existence has been in question.

In the BPS case, the construction of multi-centered attractor solutions is a simple generalization of the full attractor flows of single-centered black holes: one needs simply to replace the single-centered harmonic functions in a single-centered BPS flow with multi-centered harmonic functions.
However, the full attractor flow of a generic single-centered non-BPS black hole has not been solved analytically, due to the difficulty of solving second-order equations of motion. Ceresole et al.\ obtained an equivalent first-order equation for non-BPS attractors in terms of a ``fake superpotential,'' but the fake superpotential can only be explicitly constructed for special charges and asymptotic moduli \cite{Ceresole:2007wx, Lopes Cardoso:2007ky}. Similarly, the harmonic function procedure was only shown to apply to a special subclass of non-BPS black holes, but has not been proven for generic cases \cite{Kallosh:2006ib}.

In this talk, we will develop a method of constructing generic black hole attractor solutions, both BPS and non-BPS, single-centered as well as multi-centered, in a large class of 4D $\mathcal{N}=2$ supergravities coupled to vector-multiplets with cubic prepotentials. The method is applicable to models for which the 3D moduli spaces obtained via $c^{*}$-map are symmetric coset spaces. All attractor solutions in such a 3D moduli space can be constructed  algebraically in a unified way. Then the 3D attractor solutions are mapped back into four dimensions to give 4D extremal black holes.

The outline of the talk is as follows. Section 2 lays out the framework and presents our solution generating procedures; section 3 focuses on the theory of 4D $\mathcal{N}=2$ supergravity coupled to one vector-multiplet, and shows in detail how to determine the attractor flow generators; section 4 then uses these generators to construct single-centered attractors, both BPS and non-BPS, and proves that generic non-BPS solutions cannot be generated via the harmonic function procedure; section 5 constructs multi-centered solutions, and shows the great contrasts between BPS and non-BPS ones. We end with a discussion on various future directions.

\section{Framework}

\subsection{3D Moduli Space $\mathcal{M}_{3D}$}

The technique of studying stationary configurations of 4D supergravities by dimensionally reducing the 4D theories to 3D non-linear $\sigma$-models coupled to gravity was described in the pioneering work \cite{Breitenlohner:1987dg}. The 3D moduli space for 4D ${\cal N}=2$ supergravity coupled to $n_V$ vector-multiplets is well-studied, for example in \cite{Cecotti:1988qn,Ferrara:1989ik,de Wit:1992wf,Ceresole:1995ca}. Here we briefly review the essential points.

The bosonic part of the 4D action is:
\begin{equation}
S=-\frac{1}{16\pi}\int d^4x
\sqrt{-g^{(4)}}\left[R-2G_{i\bar{j}} dt^i\wedge \ast_4
d\bar{t}^{\bar{j}}-F^I\wedge G_I\right]
\end{equation}
where $I=0,1...n_V$, and $G_I =  (Re{\cal N})_{IJ} F^J+(Im{\cal
N})_{IJ} \ast F^J$.
For a theory endowed with a prepotential $F(X)$, $\mathcal{N}_{IJ}=F_{IJ}+2i\frac{(\textrm{Im} F\cdot
X)_I(\textrm{Im} F \cdot X)_J}{X \cdot \textrm{Im} F \cdot X}$
where $F_{IJ}=\partial_I \partial_J F(X)$ \cite{Ceresole:1995ca}.
We will consider generic stationary solutions, allowing non-zero angular momentum. The ansatz for the metric and gauge fields are:
\begin{eqnarray}
ds^2 &=&-e^{2U} (dt+\bm{\omega})^2+e^{-2U} \textbf{g}_{ab}dx^a dx^b \\
A^I&=&A^I_0(dt+\bm{\omega})+\mathbf{A}^I
\end{eqnarray}
where $\textbf{g}_{ab}$ is the 3D space metric and bold fonts
denote three-dimensional fields and operators.
The variables are $3n_V+2$ scalars $\{U,t^i,\bar{t}^{\bar{i}},A_0^I\}$, and
$n_V+2$ vectors $\{\bm{\omega}, \mathbf{A}^I\}$.

The existence of a time-like isometry allows us to reduce the 4D theory to a 3D non-linear $\sigma$-model on this isometry. Dualizing the vectors $\{\bm{\omega},\mathbf{A}^I\}$ to the
scalars $\{\sigma,B_I\}$, and renaming $A_0^I$ as $A^I$, we arrive at the 3D Lagrangian, which is a non-linear $\sigma$-model minimally coupled to 3D gravity:\footnote{
Note that the black hole potential term in 4D breaks down into kinetic terms of the 3D moduli, thus there is no potential term for the 3D moduli.
}
\begin{equation}
\mathcal{L} = \frac{1}{2}\sqrt{\textbf{g}} (
-\frac{1}{\kappa}\textbf{R}+\partial_a \phi^m
\partial^a \phi^ng_{mn})
\end{equation}
where $\phi^n$ are the $4(n_V+1)$ moduli fields $\{U,t^i,\bar{t}^{\bar{i}},\sigma,A^I,B_I\}$, and $g_{mn}$ is the
metric of the 3D moduli space $\mathcal{M}_{3D}$, whose line element is:
\begin{eqnarray}\label{metric}
ds^2&=&\textbf{d}U^2+\frac{1}{4}e^{-4U}(\textbf{d}\sigma+A^I \textbf{d}B_I-B_I
\textbf{d}A^I)^2+g_{i\bar{j}}(t,\bar{t})\textbf{d}t^i \cdot
\textbf{d}\bar{t}^{\bar{j}}\nonumber\\
&&+\frac{1}{2}e^{-2U}[(Im\mathcal{N}^{-1})^{IJ}(\textbf{d}B_I+\mathcal{N}_{IK}\textbf{d}A^K)\cdot
(\textbf{d}B_J+\overline{\mathcal{N}}_{JL}\textbf{d}A^L) ]
\end{eqnarray}
The resulting $\mathcal{M}_{3D}$ is a para-quaternionic-K\"ahler manifold, with special holonomy $Sp(2,\mathbb{R})\times
Sp(2n_V+2,\mathbb{R})$ \cite{Gunaydin:2005mx}. It is the analytical continuation of the  quaternionic-K\"ahler manifold with special holonomy $USp(2,\mathbb{R})\times
USp(2n_V+2,\mathbb{R})$ studied in \cite{Ferrara:1989ik}. Thus the vielbein has two indices $(\alpha,A)$,
transforming under $Sp(2,\mathbb{R})$ and $Sp(2n_V+2,\mathbb{R})$,
respectively. The para-quaternionic vielbein is the analytical
continuation of the quaternionic vielbein computed in
\cite{Ferrara:1989ik}. This
procedure is called the $c^{*}$-map \cite{Gunaydin:2005mx}, as it is the analytical continuation of the $c$-map in \cite{Cecotti:1988qn,Ferrara:1989ik}

The isometries of the $\mathcal{M}_{3D}$ descends from the
symmetry of the 4D system. In particular, the gauge symmetries in 4D give the
shift isometries of $\mathcal{M}_{3D}$, whose associated conserved charges are:
\begin{equation}
\label{ConservedChargesCurrents}
q_I d\tau =
J_{A^I}=P_{A^{I}}-B_{I}P_{\sigma}, \qquad p^I
d\tau=J_{B_I}= P_{B_{I}}+A^{I}P_{\sigma}, \qquad  a
d\tau=J_{\sigma}=P_{\sigma}
\end{equation}
where the $\{P_{\sigma},P_{A^I},P_{B_I}\}$ are the
momenta. Here $\tau$ is the affine parameter defined as $d\tau \equiv
-\boldsymbol{\ast}_{3}\sin{\theta}d\theta d\phi$.
$(p^I,q_I)$ are the D-brane charges, and $a$ the NUT charge. A non-zero $a$ gives rise to closed time-like
curves, so we will set $a=0$ from now on.

\subsection{Attractor Flow Equations}

The E.O.M. of $3D$ gravity is Einstein's equation:
\begin{equation}
\textbf{R}_{ab}-\frac{1}{2}\textbf{g}_{ab} \textbf{R}=\kappa
T_{ab}=\kappa (
\partial_a \phi^m
\partial_b \phi^ng_{mn}-\frac{1}{2}\textbf{g}_{ab}\partial_c \phi^m
\partial^c \phi^ng_{mn})
\end{equation}
and the E.O.M. of the 3D moduli are the geodesic equations in $\mathcal{M}_{3D}$:
\begin{equation}
\nabla_a \nabla^a \phi^n + \Gamma^{n}_{mp} \partial_a \phi^m
\partial^a
\phi^p=0
\end{equation}

It is not easy to solve a non-linear $\sigma$-model that couples to gravity. However, the theory greatly simplifies when the 3D spatial slice is flat: the dynamics of the moduli are
decoupled from that of $3D$ gravity:
\begin{equation}
T_{ab}=0=\partial_a \phi^m  \partial_b \phi^n g_{mn} \qquad \textrm{and} \qquad \partial_a \partial^a \phi^n + \Gamma^{n}_{mp} \partial_a \phi^m
\partial^a
\phi^p=0
\end{equation}
In particular, a single-centered attractor flow then corresponds to a null geodesics in $\mathcal{M}_{3D}$: $ds^2=d\phi^m d\phi^n g_{mn}=0$.

The condition of the 3D spatial slice being flat is guaranteed for BPS attractors, both single-centered and multi-centered, by supersymmetry. Furthermore, for single-centered attractors, both BPS and non-BPS, extremality condition ensures the flatness of the 3D spatial slice. In this paper, we will impose this flat 3D spatial slice condition on all multi-centered non-BPS attractors we are looking for. They correspond to the multi-centered solutions that are directly ``assembled" by single-centered attractors, and have properties similar to their single-centered constituents: they live in certain null totally geodesic sub-manifolds of $\mathcal{M}_{3D}$. We will discuss the relaxation of this condition at the end of the paper.

To summarize, the problem of finding 4D single-centered black hole attractors can be translated into finding appropriate null geodesics in $\mathcal{M}_{3D}$, and that of finding 4D multi-centered black hole bound states into finding corresponding 3D multi-centered solutions living in certain null totally geodesic sub-manifold of $\mathcal{M}_{3D}$.

The null geodesic that corresponds to a 4D black hole attractor is one that terminates at a point on the $U \rightarrow -\infty$ boundary and in the interior region with respect to all
other coordinates of the moduli space $\mathcal{M}_{3D}$. However, it is difficult to find such geodesics since a generic null geodesic flows to the boundary of $\mathcal{M}_{3D}$. For BPS attractors, the termination of the null geodesic at its attractor point is guaranteed by the constraints imposed by supersymmetry. For non-BPS attractor, one need to find the constraints without the aid of supersymmetry. We will show that this can be done for models with $\mathcal{M}_{3D}$ that are symmetric coset spaces. Moreover, the method can be easily generalized
to find the multi-centered non-BPS attractor solutions.

\subsection{Models with $\mathcal{M}_{3D}$ Being Symmetric Coset Spaces}

A homogeneous space $\mathcal{M}$ is a manifold on which its isometry group $\mathbf{G}$ acts transitively. It is isomorphic to the coset space $\mathbf{G}/\mathbf{H}$, with $\mathbf{G}$ being the isometry group and $\mathbf{H}$ the
isotropy group. For $\mathcal{M}_{3D}=\mathbf{G}/\mathbf{H}$, $\mathbf{H}$ is the maximal compact subgroup of
$\mathbf{G}$ when one compactifies on a spatial isometry down to
$(1,2)$ space, or the analytical continuation of the maximal
compact subgroup when one compactifies on the time
isometry down to $(0,3)$ space.

The Lie algebra $\mathbf{g}$ has
Cartan decomposition: $\mathbf{g}=\mathbf{h}\oplus \mathbf{k}$
where
\begin{equation}
[\mathbf{h},\mathbf{h}]=\mathbf{h} \qquad
[\mathbf{h},\mathbf{k}]=\mathbf{k}
\end{equation}
When $\mathbf{G}$ is semi-simple, the coset space $\mathbf{G}/\mathbf{H}$ is
symmetric, meaning:
\begin{equation}
[\mathbf{k},\mathbf{k}]=\mathbf{h}
\end{equation}

The building block of the non-linear $\sigma$-model with symmetric coset space $\mathcal{M}_{3D}$ as target space is the coset representative $M$, from which the left-invariant current is constructed:
\begin{equation}
J=M^{-1}dM=J_{\mathbf{k}}+J_{\mathbf{h}}
\end{equation}
where $J_{\mathbf{k}}$ is the projection of $J$ onto the coset algebra $\mathbf{k}$. The
lagrangian density of the $\sigma$-model with target space
$\mathbf{G}/\mathbf{H}$ is then given by $J_{\mathbf{k}}$ as:
\begin{equation}
\mathcal{L}=\hbox{Tr}(J_{\mathbf{k}}\wedge *_{3} J_{\mathbf{k}})
\end{equation}

The symmetric coset space has the nice property that its geodesics $M(\tau)$ are simply generated by exponentiation of the coset algebra $\mathbf{k}$:
\begin{equation}\label{Mgeodesic}
M(\tau)=M_0e^{k \tau/2}\qquad \textrm{with} \qquad k \in \mathbf{k}
\end{equation}
where $M_0$ parameterizes the initial point of the geodesic, and the factor $\frac{1}{2}$ in the exponent is for later convenience. A null geodesic corresponds to $|k|^2=0$.
Therefore, in the symmetric coset space $\mathcal{M}_{3D}$, the problem of finding the null geodesics that terminate at attractor points is translated into finding the appropriate constraints on the null elements of the coset algebra $\mathbf{k}$.

The theories with 3D moduli spaces $\mathcal{M}_{3D}$ that are symmetric coset spaces include: $D$-dimensional gravity toroidally compactified to four dimensions, all 4D ${\cal N}
> 2$ extended supergravities, and certain 4D $\mathcal{N}=2$ supergravities coupled to vector-multiplets with cubic prepotentials. The entropies in the last two classes are U-duality invariant. In this talk, we will focus on the last class. The discussion on the first class can be found in \cite{Gaiotto:2007ag}.

\subsubsection{Parametrization of $\mathcal{M}_{3D}$}

The symmetric coset space $M_{3D}=\mathbf{G}/\mathbf{H}$ can be parameterized by exponentiation of the solvable subalgebra $solv$ of $\mathbf{g}$:
\begin{equation}
 \mathcal{M}_{3D}=\mathbf{G}/\mathbf{H}=e^{solv} \qquad \textrm{with} \qquad\mathbf{g}=\mathbf{h}\oplus solv
\end{equation}
The solvable subalgebra $solv$ is determined via Iwasawa decomposition of $\mathbf{g}$. Being semi-simple, $\mathbf{g}$ has Iwasawa decomposition:
$\mathbf{g}=\mathbf{h}\oplus\mathbf{a}\oplus\mathbf{n}$, where
$\mathbf{a}$ is the maximal abelian subspace of $\mathbf{k}$, and
$\mathbf{n}$ the nilpotent subspace of the positive root space
$\Sigma^{+}$ of $\mathbf{a}$. The solvable subalgebra $solv=\mathbf{a}\oplus\mathbf{n}$. Each point $\phi^{n}$ in $\mathcal{M}_{3D}$ corresponds to a solvable element $\Sigma (\phi)=e^{solv}$, thus the solvable elements can serve as coset representatives.

We briefly explain how to extract the values of moduli from the coset representative $M$.
Since $M$ is defined up to the action of the isotropy group
$\mathbf{H}$, we need to construct from $M$ an entity that encodes the values of moduli in an $\mathbf{H}$-independent way. The symmetric matrix $S$ defined as:
\begin{equation}
S \equiv MS_0 M^T
\end{equation}
has such a property, where $S_0$ is the signature matrix.\footnote{In all systems considered in the present work, the isotropy group $\mathbf{H}$ is
the maximal orthogonal subgroup of $\mathbf{G}$: $HS_0H^T=S_0$, for any $H \in \mathbf{H}$. Therefore, $S$ is invariant under the $\mathbf{H}$-action $M\rightarrow MH$.}
Moreover, as the
isometry group $\mathbf{G}$ acts transitively on the space of matrices with signature $S_0$, the space of possible $S$ is the
same as the symmetric coset space $\mathcal{M}_{3D}=\mathbf{G}/\mathbf{H}$. Therefore, we can read off the values of moduli from $S$ in an $\mathbf{H}$-independent way.

The non-linear $\sigma$-model with target space $\mathcal{M}_{3D}$ can also be described in terms of $S$ instead of $M$. First, the left-invariant current of $S$ is $J_S=S^{-1}dS$, which is related to $J_{\mathbf{k}}$ by:
\begin{equation}
J_{S}=S^{-1}dS=2(S_0M^T)^{-1}J_{\mathbf{k}}(S_0M^T)
\end{equation}
The lagrangian density in terms of $S$ is thus $
\mathcal{L}=\frac{1}{4}\hbox{Tr}(J_{S}\wedge *_{3} J_{S})$. The equation of
motion is the conservation of current:
\begin{equation}
\nabla \cdot J=\nabla \cdot (S^{-1} \nabla S)=0
\end{equation}
where we have dropped the subscript $S$ in $J_S$, since we will
only be dealing with this current from now on.

\subsection{Example: $n_V=1$}

\begin{figure}[htbp]
  \centering
  \includegraphics{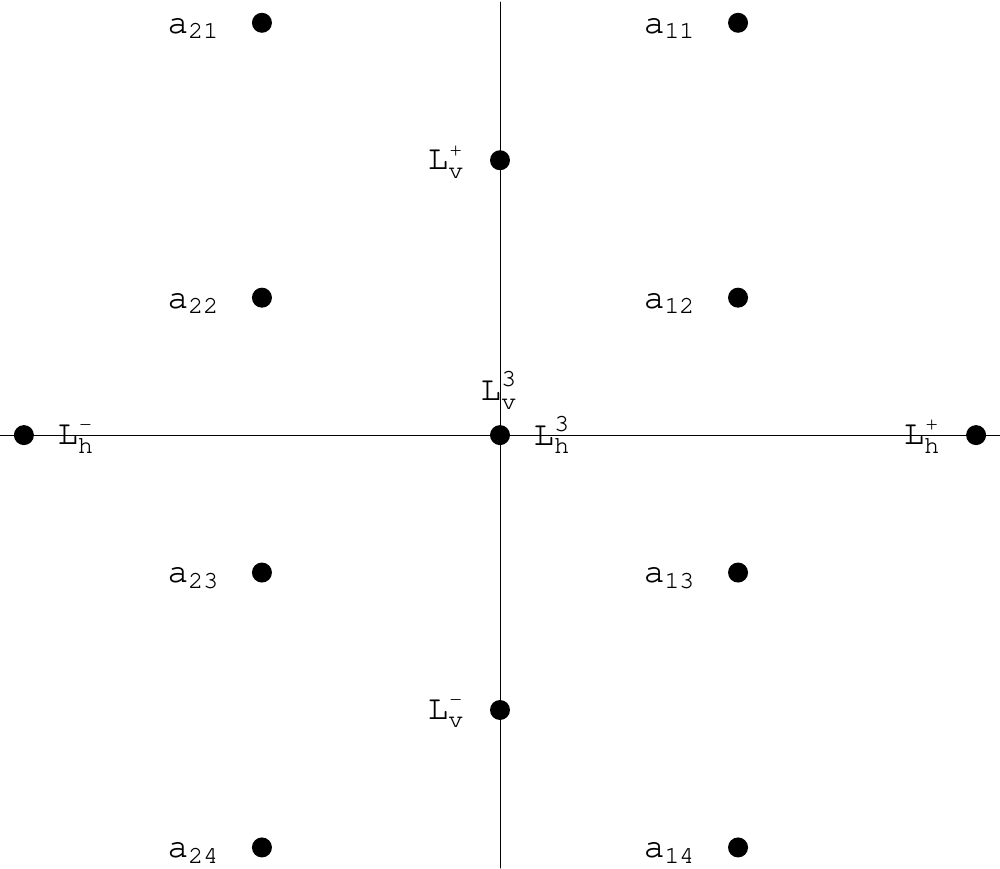}
  \caption{Root Diagram of $G_{2(2)}$ in Cartan Decomposition.}
  \label{fig:RootCartan}
\end{figure}

In this talk, we will perform the explicit computation only for the simplest case: 4D $\mathcal{N}=2$ supergravity coupled to one vector-multiplet. The generalization to generic $n_V$ is straightforward. The 3D moduli space $\mathcal{M}_{3D}$ for $n_V=1$ is an eight-dimensional quaternionic k\"ahler manifold, with special holonomy $Sp(2,\mathbb{R})\times Sp(4,\mathbb{R})$. Computing the killing symmetries of the metric (\ref{metric}) with $n_V=1$ shows that it is a coset space $G_{2(2)}/(SL(2,\mathbb{R})\times SL(2,\mathbb{R}))$ \footnote{Other work on this coset space has appeared recently, including \cite{Bouchareb:2007ax, Clement:2007qy, Gunaydin:2007qq}.}. Figure \ref{fig:RootCartan} shows the root diagram of $G_{2(2)}$ in its Cartan decomposition.
The six roots on the horizontal and vertical axes
$\{L^{\pm}_h, L^3_h, L^{\pm}_v, L^3_v\}$ generate the isotropy subgroup $\textbf{H}=SL(2,\mathbb{R})_{h}\times
SL(2,\mathbb{R})_{v}$. The two vertical columns of eight roots $a_{\alpha A}$ generate the coset algebra $\mathbf{k}$, with index $\alpha$ labeling a spin-$1/2$ representation of $SL(2,\mathbb{R})_{h}$ and index $A$ a spin-$3/2$ representation of $SL(2,\mathbb{R})_{v}$.

\begin{figure}[htbp]
  \centering
  \includegraphics{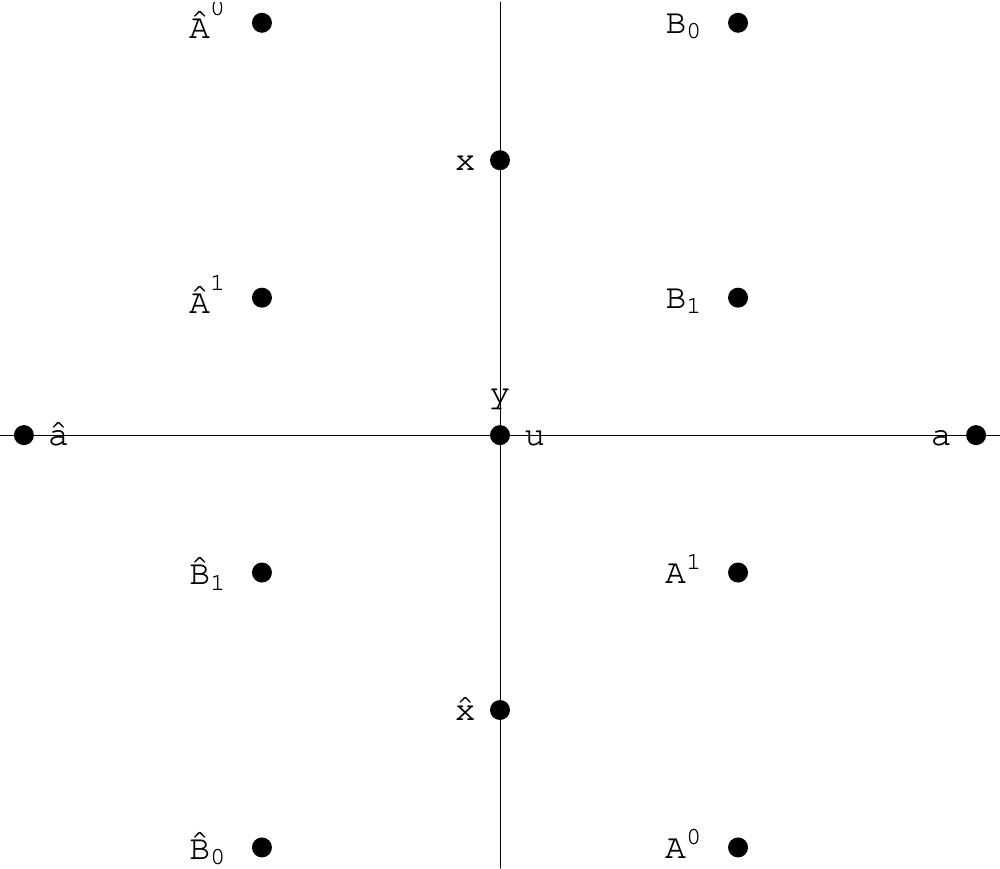}
  \caption{Root Diagram of $G_{2(2)}$ in Iwasawa Decomposition. $\{\mathbf{u},\mathbf{y},\mathbf{x},
\bm{\sigma},\mathbf{A}^{0},\mathbf{A}^{1},\mathbf{B}_{1},\mathbf{B}_{0}\}$
generates the solvable subgroup $Solv$.}
  \label{fig:RootSolv}
\end{figure}

The Iwasawa decomposition, $\mathbf{g}=\mathbf{h}\oplus solv$ with $solv=\mathbf{a}\oplus\mathbf{n}$, is shown in Figure \ref{fig:RootSolv}. 
The two Cartan generators $\{\mathbf{u},
\mathbf{y}\}$ form $\mathbf{a}$, while $\mathbf{n}$ is spanned by $\{\mathbf{x},
\bm{\sigma},\mathbf{A}^{0},\mathbf{A}^{1},\mathbf{B}_{1},\mathbf{B}_{0}\}$. $\{\mathbf{u},\mathbf{y}\}$ generates the
rescaling of $\{u,y\}$, where $u\equiv e^{2U}$, and $\{\mathbf{x},
\bm{\sigma},\mathbf{A}^{0},\mathbf{A}^{1},\mathbf{B}_{1},\mathbf{B}_{0}\}$
generates the translation of $\{x,\sigma,A^0,A^1,B_1,B_0\}$ \cite{de Wit:1992wf}.

The moduli space $\mathcal{M}_{3D}$ can be parameterized by solvable elements:
\begin{equation}
\Sigma(\phi)=e^{(\ln{u})\mathbf{u}/2+(\ln{y})\mathbf{y}}e^{x\mathbf{x}+A^I\mathbf{A}^I+B_I\mathbf{B_I}+\sigma\bm{\sigma}}
\end{equation}
The symmetric matrix $S$ can then be expressed in terms of the
eight moduli $\phi^n$:
\begin{equation}
S(\phi)=\Sigma(\phi)S_0\Sigma(\phi)^T
\end{equation}
which shows how to extract the values of moduli from $S$ even when $S$ is not constructed from the solvable elements, since it is invariant under $\mathbf{H}$-action.

\section{Generators of Attractor Flows}

In this section, we will solve 3D attractor flow generators $k$ as in (\ref{Mgeodesic}). We will prove that extremality condition ensures that they are nilpotent elements of the coset algebra $\mathbf{k}$. In particular, for $n_V=1$, both BPS and non-BPS generators are third-degree nilpotent. However, despite this common feature, $k_{BPS}$ and $k_{NB}$ differ in many aspects.

\subsection{Construction of Attractor Flow Generators}
\subsubsection{Construction of $k_{BPS}$}

Since the 4D BPS attractor solutions are already known, one can easily obtain the BPS flow generator $k_{BPS}$ in the 3D moduli space $\mathcal{M}_{3D}$.

The generator $k_{BPS}$ can be expanded by coset elements $a_{\alpha A}$ as $k_{BPS}=a_{\alpha A} C^{\alpha A} $, where $C^{\alpha A}$ are conserved along the flow.
On the other hand, since the conserved currents in the homogeneous space are
constructed by projecting the one-form valued Lie algebra
$g^{-1}\cdot dg$ onto $\mathbf{k}$, a procedure that also gives the vielbein: $J_{\mathbf{k}}=g^{-1} dg|_{\mathbf{k}} =  a_{\alpha A}V^{\alpha A}$, the vielbein $V^{\alpha A}$ are also conserved along the flow: $\frac{d}{d\tau}\left(V_n^{\alpha A} \dot{\phi}^n \right) = 0$.
Since both the expansion coefficients $C^{\alpha A}$ and the vielbein $V^{\alpha A}$ transform as $\mathbf{(2,4)}$ of $SL(2,\mathbb{R})_h \times (SL(2,\mathbb{R})_v$ and are conserved along the flow, they are related by:
\begin{equation}\label{CVBPS}
C^{\alpha A} = V_n^{\alpha A} \dot{\phi}^n
\end{equation}
up to an overall scaling factor.

In terms of the vielbein $V^{\alpha A}$, the supersymmetry condition that gives
the BPS attractors is: $V^{\alpha A}=z^{\alpha}V^{A}$ \cite{Gunaydin:2005mx,Pioline:2006ni,Gunaydin:2007qq}.  Using (\ref{CVBPS}), we conclude that the 3D BPS flow generator $k_{BPS}$ has the expansion
\begin{equation}
k_{BPS}=a_{\alpha A}z^{\alpha}C^A
\end{equation}

A 4D supersymmetric black hole is labeled by four D-brane charges $(p^0,p^1,q_1,q_0)$. A 3D attractor flow generator $k_{BPS}$ has five parameters $\{C^{A},z\}$. As
will be shown later, $z$ drops off in the final solutions of BPS attractor flows, under the zero NUT charge condition. Thus the geodesics generated by $k_{BPS}$ are indeed in a four-parameter family.

$k_{BPS}$ can be obtained by a twisting procedure as follows.  First, define a $k^0_{BPS}$ which is spanned by the four coset generators with positive charges under $SL(2,\mathbb{R})_{h}$:
\begin{equation}
k^0_{BPS} \equiv a_{1A}C^A
\end{equation}
then, conjugate $k^{0}_{BPS}$ with lowering operator $L^{-}_{h}$:
\begin{equation}\label{twistingBPS}
k_{BPS}=e^{-zL^{-}_{h}}k^0_{BPS}e^{zL^{-}_{h}}
\end{equation}

Using properties of $k^0_{BPS}$, it is easy to check that $k_{BPS}$
is null:
\begin{equation}
|k_{BPS}|^2=0
\end{equation}
More importantly, $k_{BPS}$ is found to be third-degree nilpotent:
\begin{equation}
k_{BPS}^3=0
\end{equation}
 A natural question then arises: Is the nilpotency condition of $k_{BPS}$ a result of supersymmetry or extremality? If latter, we can use the nilpotency condition as a constraint to solve for the non-BPS attractor generators $k_{NB}$. We will prove that this is indeed the case.

\subsubsection{Extremality implies nilpotency of flow generators}

We will now prove that all attractor flow generators, both BPS and Non-BPS, are nilpotent elements in the coset algebra $\mathbf{k}$. It is a result of the near-horizon geometry of extremal black holes.

The near-horizon geometry of a $4D$ attractor is $AdS_2 \times
S^2$, i.e.
\begin{equation}
e^{-U}\rightarrow \sqrt{V_{BH}}|_{*} \tau  \qquad \textrm{as}
\qquad \tau \rightarrow \infty
\end{equation}
As the flow goes to the near-horizon, i.e., as $u=e^{2U} \rightarrow 0$, the solvable element $M=e^{(\ln{u})\mathbf{u}/2+\cdots}$ is a polynomial function of $\tau$:
\begin{equation}
M(\tau)\sim u^{-\ell/2}\sim \tau^{\ell}
\end{equation}
where $-\ell$ is the lowest eigenvalue of $\mathbf{u}$.

On the other hand, since the geodesic flow is generated by $k \in \mathbf{k}$ via $M(\tau)=M_0e^{k \tau/2}$, $M(\tau)$ is an exponential function of $\tau$. To reconcile
the two statements, the attractor flow generator $k$ must be nilpotent:
\begin{equation}
k^{\ell+1}=0
\end{equation}
where the value of $\ell$ depends on the particular moduli space under consideration. In $G_{2(2)}/SL(2,\mathbb{R})^2$, by looking at the weights of the fundamental representation, we see that $\ell=2$, thus
\begin{equation}
k^{3}=0
\end{equation}

The nilpotency condition of the flow generators also automatically
guarantees that they are null:
\begin{equation}
k^3=0 \qquad \Longrightarrow \qquad (k^2)^2=0 \qquad \Longrightarrow \qquad
\textrm{Tr}(k^2)=0
\end{equation}

\subsubsection{Construction of $k_{NB}$}
To construct non-BPS attractor flows, one needs to find third-degree nilpotent
elements in the coset algebra $\mathbf{k}$ that are distinct from the BPS ones.
In the real $G_{2(2)}/SL(2,\mathbb{R})^2$, there are two third-degree
nilpotent orbits in total \cite{Collingwood:1993}. We have shown that $k_{BPS}=e^{-zL^{-}_{h}}k^0_{BPS}e^{zL^{-}_{h}}$, with $k^{0}_{BPS}$ spanned by the four generators with positive charge under $SL(2,\mathbb{R})_h$.

Since there are only two $SL(2,\mathbb{R})$'s inside $\mathbf{H}$, a natural guess for $k_{NB}$ is that it can be constructed by the same twisting procedure with $SL(2,\mathbb{R})_h$ replaced by $SL(2,\mathbb{R})_v$:
\begin{equation}\label{knonBPSdefine}
k_{NB}=e^{-zL_{v}^{-}}k^0_{NB}e^{zL_{v}^{-}} \qquad \textrm{with} \qquad k^0_{NB}\equiv a_{\alpha a}C^{\alpha a}, \qquad \alpha, a=1,2
\end{equation}
where $k^{0}_{NB}$ is spanned by the four generators with positive charge under $SL(2,\mathbb{R})_v$.

Using properties of $k^{0}_{NB}$, one can easily show that $k_{NB}$ defined above is indeed third-degree nilpotent:
\begin{equation}
k_{NB}^3=0
\end{equation}
That is, $k_{NB}$ defined in (\ref{knonBPSdefine}) generates non-BPS attractor flows in $\mathcal{M}_{3D}$.

A 4D non-BPS extremal black hole is labeled by four D-brane charges $(p^0,p^1,q_1,q_0)$. Similar to the BPS case, the 3D attractor flow generator $k_{NB}$ has five parameters $\{C^{\alpha a},z\}$. As will be shown later, $z$ can be determined in terms of $\{C^{\alpha a}\}$ using the zero NUT charge condition, thus the geodesics generated by $k_{NB}$ are also in a four-parameter family.

\subsection{Properties of Attractor Flow Generators}

We choose the representation of $G_{2(2)}$ group to be the symmetric $7 \times
7$ matrices that preserve a
non-degenerate three-form $w_{ijk}$ such that
$\eta_{is}\equiv w_{ijk}w_{stu}w_{mno} \epsilon^{jktumno}$ is a metric
with signature $(4,3)$ and normalized as $\eta^2=1$. We
decompose $\mathbf{7}$ as $\mathbf{3} \oplus \bar{\mathbf{3}} \oplus \mathbf{1}$ of $SL(3,\mathbb{R})$ and choose the
non-zero components of $w$, $\mathbf{3} \wedge \mathbf{3} \wedge \mathbf{3}$, $\bar{\mathbf{3}}
\wedge \bar{\mathbf{3}} \wedge \bar{\mathbf{3}}$ and $\mathbf{3} \otimes \bar{ \mathbf{3}} \otimes \mathbf{1}$, as
\begin{equation}
w=dx_1 \wedge dx_2 \wedge dx_3 + dy^1 \wedge dy^2 \wedge dy^3 - \frac{1}{\sqrt{2}} dx_a \wedge dy^a \wedge dz
\end{equation}
which gives $\eta= dx_ady^a - dz^2$.
Written explicitly, an element of $G_{2(2)}$ Lie algebra is
\begin{equation}
\label{kBasis} g = \left(\begin{array}{ccc} A_{i_1}^{j_1} &
\epsilon_{i_1 j_2 k} v^k & \sqrt{2} w_{i_1} \\ \epsilon^{i_2 j_1 k} w_k
& -A_{j_2}^{i_2} & -\sqrt{2} v^{i_2} \\ -\sqrt{2} v^{j_1} & \sqrt{2}
w_{j_2} & 0
\end{array} \right)
\end{equation}
Here $A$ is a traceless $3\times 3$ matrix. The signature matrix $S_0$ is thus $\textrm{Diag}[1,-1,-1,1,-1,-1,1]$.

The real $G_{2(2)}$ group has two third-degree nilpotent orbits. In both orbits, $k^2$ is of rank two and has Jordan form with two blocks of size 3. Thus $k^2$ can be written as
\begin{equation}\label{kSquaredNullVectors}
k^2=\sum_{a,b=1,2} v_a v_b^T c_{ab}S_0
\end{equation}
with $v_a$ null and orthogonal to each other: $v_a \cdot v_b\equiv
v^T_aS_0v_b=0$, and $c_{ab}$ depends on the particular choice of
$k$. Therefore, $k$ can be expressed as:
\begin{equation}
\label{kInTermsOfvw} k=\sum_{a=1,2} (v_a w_a^T+ w_a v_a^T)S_0
\end{equation}
where each $w_a$ is orthogonal to both $v_a$: $w_a \cdot v_b=0$,
and $w_{a}$ satisfies $w_a \cdot w_b=c_{ab}$. Next we solve for $v_a$ and $w_a$ for $k_{BPS}$ and $k_{NB}$ and compare their properties.

\subsubsection{Properties of $k_{BPS}$}

The null space of $k^2$ is five-dimensional, with $v_a$ spanning its two-dimensional complement. For $k_{BPS}$, $v^{BPS}_a$ and $w^{BPS}_a$ in (\ref{kInTermsOfvw}) are solved in
terms of $C^A$ and $z$.

In basis (\ref{kBasis}), from inspection of $k_{BPS}^2$, we find that
$v^{BPS}_a$ can always be chosen to have the form:\footnote{There are some freedom on the choice of
$(v_a, w_a)$: a rotational freedom: $(v_a,w_a)\rightarrow (R_{ab}v_b,R_{ab}w_b)$ with $R$ orthogonal; and a rescaling freedom: $(v_a,w_a) \rightarrow ( v_a r,w_a/r)$.}
\begin{equation}
v^{BPS}_1=(V_1,-\eta_1V_1,0) \qquad v^{BPS}_2=(-V_2,\eta_1 V_2,\sqrt{2})
\end{equation}
where $\eta_1$ is a 3D signature matrix $\eta=\textrm{Diag}[1,-1,-1]$, and
$V_a$ are two three-vectors satisfying
\begin{equation}\label{V12condition}
V_1\cdot V_1=0 \qquad V_1\cdot V_2=0 \qquad V_2\cdot V_2=-1
\end{equation}
We drop the superscript ``BPS" for $V_a$ here since, as will be shown later, $v^{NB}_a$ can also be written in terms of $V_a$, though in a slightly different form. Note that for $k_{BPS}$, $V_2$ is defined up to a shift of $V_1$: $V_2\rightarrow V_2-c V_1$, since any linear combination of $v^{BPS}_a$ forms a new set of $v^{BPS}_a$.

Written in twistor representation,\footnote{With the inner product of two
three-vectors defined as $
V_a\cdot V_b \equiv V^T_a \eta_1 V_b$, the twistor representation of a three-vector $V=(x,y,z)$ can be chosen as
\begin{displaymath}
\sigma_{V}=x \sigma_0+y\sigma_3+z\sigma_1=\left(\begin{array}{cc} x+y&z\\
z&x-y\end{array} \right)
\end{displaymath}} $V_a$ are given by
the twistors $z$ and $u$ as
\begin{equation}
\label{ABSpinorForm} V_1^{\alpha\beta}=2z^{\alpha}z^{\beta} \qquad
\qquad V_2^{\alpha \beta} = z^{\alpha} u^{\beta} + z^{\beta}
u^{\alpha}
\end{equation}
where we have used the rescaling freedom to set $z^1u^2-z^2u^1=1$. Note that for $k_{BPS}$, the twistor $u$ is arbitrary, due to the shift freedom of $V_2$.

The condition $w^{BPS}_{a}\cdot v^{BPS}_{b}=0$ dictates that $w^{BPS}_a$ has the form:
\begin{equation}\label{W12formBPS}
w^{BPS}_1 = (W^{BPS}_1,\eta_1 W^{BPS}_1,0) \qquad w^{BPS}_2 = (W^{BPS}_2, \eta_1 W^{BPS}_2,0)
\end{equation}
with $W^{BPS}_a$ solved as:
\begin{equation}\label{W1W24D}
(W^{BPS}_1,W^{BPS}_2)^{\alpha\beta}=(P^{\alpha\beta\gamma}u_{\gamma},P^{\alpha\beta\gamma}z_{\gamma})
\end{equation}
where the totally symmetric $P^{\alpha\beta\gamma}$ is defined in terms of ${C^A}$ as
\begin{equation}
P^{111}=C^1 \qquad P^{112}=C^2 \qquad P^{122}=C^3 \qquad
P^{222}=C^4
\end{equation}
In summary, $v^{BPS}_a$ span a one-dimensional space (since $u$ is arbitrary) and $w^{BPS}_a$ span a four-dimensional space.

\subsection{Properties of $k_{NB}$}

$(v^{NB}_a,w^{NB}_a)$ are solved in terms of $\{C^{\alpha a},z\}$. The forms of $v^{NB}_a$ are only slightly different from those of $v^{BPS}_{a}$:
\begin{equation}
v^{NB}_1 = (V_1, \eta_1 V_1,0)  \qquad v^{NB}_2 = (V_2,-\eta_1 V_2,\sqrt{2})
\end{equation}
where $V_a$ are the same three-vectors given in (\ref{ABSpinorForm}), with one
major difference: the twistor $u$ is no longer arbitrary, but is determined by
$C^{\alpha a}$ as
\begin{equation}\label{NonBPSu}
u=\frac{u^2}{u^1}=\frac{C^{22}}{C^{12}}
\end{equation}
since the $V_2$ in $v^{NB}_a$ no longer has the shift freedom.

The forms of $w^{NB}_a$ are also only slightly different from the BPS
ones (\ref{W12formBPS}):
\begin{equation}\label{W12formNB}
w^{NB}_1 = (W^{NB}_1,- \eta_1 W^{NB}_1,0) \qquad w^{NB}_2 = (W^{NB}_2, \eta_1 W^{NB}_2,0)
\end{equation}
with $W^{NB}_a$ solved in terms of $\{C^{\alpha a},z,u\}$ as:
\begin{eqnarray}
(W^{NB}_1)^{\alpha\beta}&=&u^{\alpha}u^{\beta}+(C^{11}u^2-C^{12}u^1)z^{\alpha}z^{\beta}\\
(W^{NB}_2)^{\alpha\beta}&=&(z^{\alpha}u^{\beta}+u^{\alpha}z^{\beta})+(C^{21}-C^{11}z-3u^1)z^{\alpha}z^{\beta}
\end{eqnarray}
Since the value of $u$ imposes an extra constraint
on the vectors $w^{NB}_a$ via (\ref{NonBPSu}), $w^{NB}_a$ span a three-dimensional space instead of a four-dimensional one as in the BPS case (\ref{W1W24D}).
In summary, in contrast to the BPS case, $v^{NB}_a$ span a two-dimensional space and $w^{NB}_a$ span a three-dimensional one.

\section{Single-centered Attractor Flows}

Having solved the attractor flow generators for both BPS and non-BPS case, we are ready to construct single-centered attractor flows. A geodesic starting from arbitrary asymptotic moduli is given by $M(\tau)=M_0e^{k\tau/2}$, which gives the flow of $S$ as
$S(\tau)=M_0 e^{k\tau}S_0 M^T_0$, which in turn can be written as
$S(\tau)=e^{K(\tau)}S_0$, where $K(\tau)$ is a matrix function.
From now on, we use capital $K$ to denote the matrix
function which we exponentiate to generate attractor solutions.

The current of $S$ is
\begin{equation}
J =S^{-1} \nabla S=S_0\left(\nabla K + [\nabla
K,K]+\frac{1}{2}[[\nabla
K,K],K]+\cdots\right)S_0
\end{equation}
The equation of motion is the conservation of currents: $
\nabla \cdot (S^{-1} \nabla S)=0$, which is solved by $K(\tau)$ being harmonic:
\begin{equation}
\nabla^2 K(\tau)=0 \qquad \Longrightarrow \qquad K(\tau)=k\tau +g
\end{equation}

$g$ parameterizes the asymptotic moduli. Using the $\mathbf{H}$-action, we can adjust $g$ such that $g\in \mathbf{k}$, and $g$ has the same properties as the flow generator $k$, namely, $g^3=0$ and $g^2$ is of rank two. Therefore, for single-centered flow given by $S(\tau)=e^{K(\tau)}S_0$, the harmonic matrix function $K(\tau)$ has the same properties as the flow generator $k$:
\begin{equation}\label{constraintonK}
K^3(\tau)=0 \qquad \textrm{and} \qquad K^2(\tau) \,\,\textrm{rank
two}
\end{equation}

To find the harmonic $K(\tau)$ that satisfies the constraints (\ref{constraintonK}), recall that the constraints dictate $K(\tau)$ to have the form:
\begin{equation}
K(\tau)=\sum_{a=1,2}(  v_a(\tau) w_a(\tau)^T  +  w_a(\tau) v_a(\tau)^T)S_0
\end{equation}
with $v_a(\tau)$ being null and $w_a(\tau)$ orthogonal to $v_b(\tau)$ for all $\tau$. Then the constraints (\ref{constraintonK}) can simply be solved by choosing $v_a(\tau)$ to be the constant null vectors $v_a(\tau)=v_a$ and $w_a(\tau)$ to be harmonic vectors which are everywhere orthogonal to $v_b$:
\begin{equation}
w_a(\tau)=w_a\tau+m_a \qquad \textrm{with} \qquad w_a\cdot v_b=m_a\cdot v_b=0
\end{equation}
The two 7-vectors $w_a$'s contain the information of the black hole charges, and the two 7-vectors $m_a$'s contain that of asymptotic moduli.

To summarize, the single-centered attractor flow starting from an arbitrary asymptotic moduli is generated by $S(\tau)=e^{K(\tau)}S_0$, with harmonic matrix function $K(\tau)=k\tau+g$ where
\begin{equation}\label{kgB}
k=\sum_{a=1,2} [v_a w_a^T+ w_a v_a^T]S_0  \qquad \textrm{and}
\qquad g=\sum_{a=1,2}  [v_a m_a^T+ m_a v_a^T]S_0
\end{equation}
Since $k$ and $g$ share the same set of null vectors $v_a$ and both $w_a$ and $m_a$ are orthogonal to $v_b$, $g$ has the same form as that of flow generator $k$, namely:
\begin{equation}
g_{BPS}=a_{\alpha A}z^{\alpha}G^{A} \qquad \qquad g_{NB}=e^{-zL^{-}_{v}}(a_{\alpha a}G^{\alpha a})e^{zL^{-}_{v}}
\end{equation}
which guarantees that $g$ is also third-degree nilpotent.
Moreover, that $g$ and $k$ have the same form implies $[[k,g],g]=0$, thus the current
is reduced to
\begin{equation}\label{current}
J=\frac{S_0(k+\frac{1}{2}[k,g])S_0}{r^2}\hat{\vec{r}}
\end{equation}
from which we can solve $v_a$ and $w_a$ in terms of charges and asymptotic moduli.

Now that we are able to construct arbitrary attractor flows in the 3D moduli space, we can lift them to the 4D black hole attractor solution. First, in representation given by (\ref{kBasis}), the 4D moduli $t=x+iy$ can be extracted from the symmetric
matrix $S$ via:
\begin{equation}\label{xyfromS}
x(\tau)=-\frac{S_{35}(\tau)}{S_{33}(\tau)} \qquad
y=\sqrt{\frac{S_{33}(\tau)S_{55}(\tau)-S_{35}(\tau)^2}{S_{33}(\tau)^2}}
\end{equation}
and $u=e^{2U}$ via:
\begin{equation}\label{ufromS}
u=\frac{1}{\sqrt{S_{33}(\tau)S_{55}(\tau)-S_{35}(\tau)^2}}
\end{equation}
Since both $k$ and $g$ are third-degree nilpotent, $S(\tau)$ is a quadratic function of $\tau$. Moreover, since $g$ has the same form as $k$, $S(\tau)$ is composed of harmonic functions of $\tau$: $H^A(\tau)\equiv C^A \tau+G^A$ for BPS attractors and $H^{\alpha a}(\tau)\equiv C^{\alpha a} \tau+G^{\alpha a}$ for non-BPS attractors.\footnote{Space prohibits listing the rather lengthy result of $S(\tau)$, readers can consult eq. (6.2) and eq. (6.3) of \cite{Gaiotto:2007ag} for BPS attractors, and eq. (6.28) for non-BPS ones.} Generic single-centered attractor flows with arbitrary charges and asymptotic moduli can thus be generated. The attractor moduli are read off from $S(\tau)$ with $\tau \rightarrow \infty$, and asymptotic moduli with $\tau\rightarrow 0$.

The D-brane charges can be read off from the charge matrix defined as $\mathbf{Q}\equiv \frac{1}{4\pi}\int \nabla \cdot J$. The $4D$ gauge currents sit in the current $J=S^{-1}\nabla S$ as:
\begin{equation}
\label{ExtractCharges}
(J_{31},J_{51},J_{72},J_{12},J_{32})=(\sqrt{2} J_{B_0},-\sqrt{2} J_{B_1},\frac{2}{3}J_{A^1} ,\sqrt{2}J_{A^0},-2 J_{\sigma})
\end{equation}
Therefore $\mathbf{Q}$ relates to the D-brane charge $(p^0,p^1,q_1,q_0)$ and the vanishing NUT charge $a$ by
\begin{equation}\label{QDbranes}
(\mathbf{Q}_{31},\mathbf{Q}_{51},\mathbf{Q}_{72},\mathbf{Q}_{12})=(\sqrt{2}p^0,-\sqrt{2}p^1,\frac{2}{3}q_1,\sqrt{2}q_0)
\qquad \mathbf{Q}_{32}=-2a=0
\end{equation}

\subsection{Single-Centered BPS Attractor Flows}
\begin{figure}[htbp]
  \centering
  \includegraphics{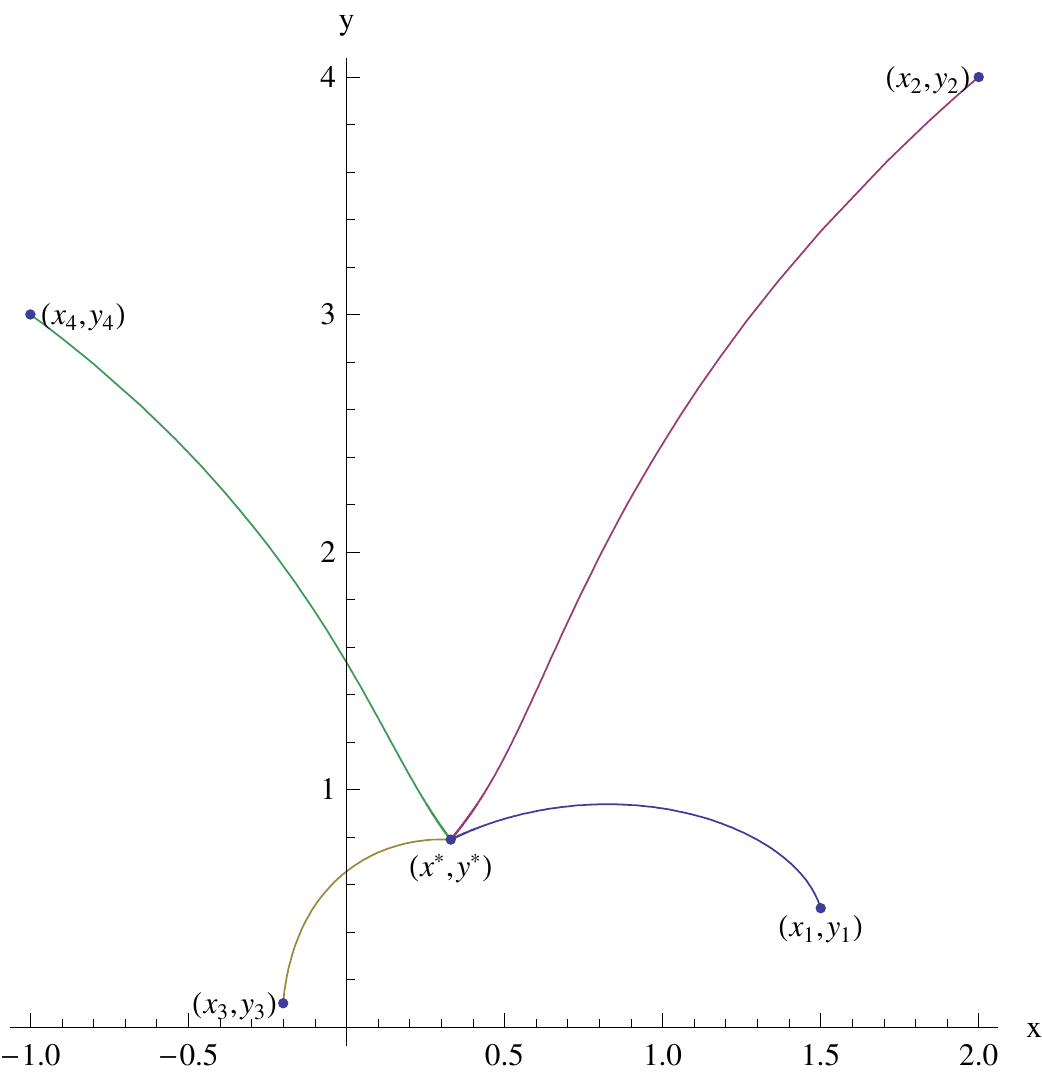}
  \caption{BPS flow with charge $(p^0, p^1, q_1, q_0)= (5, 2,7,-3)$ and attractor point $(x^*,y^*)=(0.329787,0.788503)$. The initial points of each flow are given by  $(x_1=1.5, y_1=0.5), (x_2=2, y_2=4), (x_3=-0.2, y_3=0.1), (x_4=-1, y_4=3)$.}
  \label{fig:BPSflow}
\end{figure}

  As an example, a single-centered BPS black hole constructed by lifting the attractor solution in $\mathcal{M}_{3D}$ is shown in Figure \ref{fig:BPSflow}.
It has D-brane charges $(p^0, p^1, q_1, q_0)= (5, 2,7,-3)$.
The four flows, starting from different asymptotic moduli, terminate at the attractor point $(x_{BPS}^*,y_{BPS}^*)$ with different tangent directions. The reason is that the mass matrix of the black hole potential $V_{BH}$ at the BPS critical point has two identical eigenvalues, thus there is no preferred direction for the geodesics to flow to the attractor point.

We now discuss in detail how to determine $k_{BPS}$ and $g_{BPS}$ for given charges and asymptotic moduli. There are nine parameters in $k_{BPS}$ and $g_{BPS}$: $\{C^A, G^A,z\}$,
since the twistor $u$ is arbitrary. On the other hand, there are eight constraints in a given attractor flow: four
D-brane charges $(p^I,q_I)$, the vanishing NUT charge $a$, and the asymptotic moduli
$(x_0, y_0,u_0)$.\footnote{The asymptotic value of $u$ can be fixed to an arbitrary value by a rescaling of time and the radial distance. We will set $u_0=1$.}
We will use these eight constraints to fix $C^A$ and $G^A$ in $k_{BPS}$ and $g_{BPS}$, leaving the twistor $z$ unfixed.

Integrating the current (\ref{current}) for BPS case produces five coupled equations:
\begin{equation}\label{kfromQsingleg}
\mathbf{Q}_{BPS}=S_0 (k_{BPS}+\frac{1}{2}[k_{BPS}, g_{BPS}])S_0
\end{equation}
where $[k_{BPS},g_{BPS}]=\langle C,G \rangle \Theta$, with $\langle C,G \rangle \equiv C^1
G^4-3C^2 G^3+3C^3 G^2-C^4 G^1$, and  $\Theta \equiv
-\frac{4}{1+z^2}e^{-zL^{-}_{h}}L^{+}_he^{zL^{-}_{h}}$.

In order to show that the BPS flow can be expressed in terms of
harmonic functions: $H(\tau)=Q\tau+h$, with $Q\equiv(p^I, q_I)$ and $h\equiv(h^I, h_I)$,
we will solve $g_{BPS}$ in terms of $h$ instead of $(x_0, y_0,u_0)$. $h$ relates to the asymptotic moduli by
\begin{equation}\label{hconstraintxyu}
(x_0,y_0,u_0)_{BPS}=(x,y,u)_{BPS}^{*}(Q\rightarrow h)
\end{equation}
and there is one extra degree of freedom to be fixed later.

First, for later convenience, we separate from $g_{BPS}$ a piece that has the same dependence on $(h,z)$ as $k_{BPS}$ on $(Q,z)$:
\begin{equation}\label{gghlambda}
g_{BPS}=g_{BPS,h}+\Lambda \qquad \textrm{with} \qquad g_{BPS,h}\equiv k_{BPS}(Q\rightarrow h, z)
\end{equation}
that is, $g_{BPS,h}=a_{\alpha A}z^{\alpha}G^{A}_{h}$ with $G^{A}_{h}\equiv C^{A}(Q\rightarrow h)$. We can use the unfixed degree of freedom in $h$ to set $\langle C,G_h \rangle=0$, so that (\ref{kfromQsingleg}) simplifies into
\begin{equation}\label{kfromQsinglelambda}
\mathbf{Q}_{BPS}=S_0 (k_{BPS} +\frac{1}{2}[k_{BPS}, \Lambda])S_0
\end{equation}
$\Lambda$ can then be determined using the three constraints from (\ref{hconstraintxyu}) and the zero NUT charge condition in (\ref{kfromQsinglelambda}): $\Lambda=a_{\alpha A}z^{\alpha}E^A $ with
$E^1=-E^3=-\frac{1}{1+z^2}$ and $E^2=-E^4=\frac{z}{1+z^2}$. The form of $\Lambda$ will ensure that the twistor $z$ drops off in the final attractor flow solution written in terms of $Q$ and $h$.

The remaining four conditions in the coupled equations
(\ref{kfromQsinglelambda}) determine $C^A$ as functions of D-brane charges and the twistor $z$: $C^{A}=C^{A}(Q,z)$.\footnote{See eq. (6.18) of \cite{Gaiotto:2007ag} for the full solutions.}
Then $G_h^A$ are given by $G^A_h=C^A(Q\rightarrow h,z)$. The product $\langle C^A, G^A_h \rangle$ is
proportional to the symplectic product of $(p^I,q_I)$ and $(h^I,
h_I)$:
\begin{equation}
\langle C^A, G^A_h \rangle=\frac{2}{1+z^2}<Q,h>\qquad
\textrm{where} \qquad <Q,h>\equiv p^0h_0+p^1h_1-q_1h^1-q_0h^0
\end{equation}
The condition $\langle C^A, G^A_h \rangle=0$ is then the
integrability condition on $h$: $<Q,h>=0$.

BPS attractor flows in terms
of $(p^I, q_I)$ and $(h^I, h_I)$ are obtained by substituting solutions of $C^A(Q,z)$ and $G^A_h(h,z)$ into the flow of $S(\tau)$. The attractor moduli are determined by the charges as:
\begin{equation}\label{moduli*QB}
x^{*}_{BPS} = -\frac{p^0 q_0+p^1 \frac{q_1}{3}}{2[(p^1)^2+p^0
\frac{q_1}{3}]} \qquad y^*_{BPS} =
\frac{\sqrt{J_4(p^0,p^1,\frac{q_1}{3}, q_0)}}{2[(p^1)^2+p^0
\frac{q_1}{3}]}
\end{equation}
where $J_4(p^0,p^1,q_1, q_0)$ is the quartic $E_{7(7)}$ invariant:
\begin{equation}
J_4(p^0,p^1,q_1, q_0)=3 (p^1q_1)^2 -6(p^0q_0) (p^1
q_1)-(p^0q_0)^2-4(p^1)^3 q_0+ 4 p^0(q_1)^3
\end{equation}
thus $J_4(p^0,p^1,\frac{q_1}{3}, q_0)$ is the discriminant of charges. Charges with positive (negative) $J_4(p^0,p^1,\frac{q_1}{3}, q_0)$ form a BPS (non-BPS) black hole. The attractor value of $u$ is $u^*_{BPS} = 1/\sqrt{J_4(p^0,p^1,\frac{q_1}{3}, q_0)}$.
The constraint on $h$ from $u_0=1$ is then
$J_4(h^0,h^1,\frac{h_1}{3},h_0)=1$. The attractor moduli (\ref{moduli*QB}) match those from Type II string compactified on diagonal $T^6$, with $q_1 \rightarrow \frac{q_1}{3}$.

Now we will prove that the BPS attractor flows constructed above can indeed be
generated by the ``naive" harmonic function procedure, namely, by replacing charges $Q$ in the attractor moduli with the corresponding harmonic functions $Q\tau+h$. First, using
the properties of $\Lambda$, the flow of $t=x+iy$ can be generated from the attractor
moduli by replacing $k_{BPS}$ with the harmonic function $k_{BPS}\tau+g_{BPS,h}$:
\begin{equation}\label{kharmonic}
t_{BPS}(\tau)=t_{BPS}^{*}(k_{BPS}\rightarrow k\tau+g_{BPS,h})
\end{equation}
Then, since $k_{BPS}$ and $g_{BPS,h}$ share the same twistor $z$, this is equivalent
to replacing $C^A$ with harmonic functions $H^A(\tau)=C^A\tau+G^A_h$
while keeping the twistor $z$ fixed:
\begin{equation}\label{Charmonic}
t_{BPS}(\tau)= t_{BPS}^{*}(C^A\rightarrow C^A\tau+G^A_h,z)
\end{equation}
Finally, since $C^A$ is linear in $Q$ and $G^A_h$ linear in $h$, and
since $z$ drops off after plugging in the solutions $C^A(Q,z)$ and $G^A_h(h,z)$, we conclude that the flow of $t_{BPS}(\tau)$ is given by replacing the charges $Q$ in the
attractor moduli with the corresponding harmonic functions $Q\tau+h$:
\begin{equation}\label{QharmonincProcedure}
t_{BPS}(\tau)=t^{*}_{BPS}(Q\rightarrow Q\tau+h)
\end{equation}

\subsection{Single-Centered Non-BPS Attractor Flows}

\begin{figure}[htbp]
  \centering
  \includegraphics{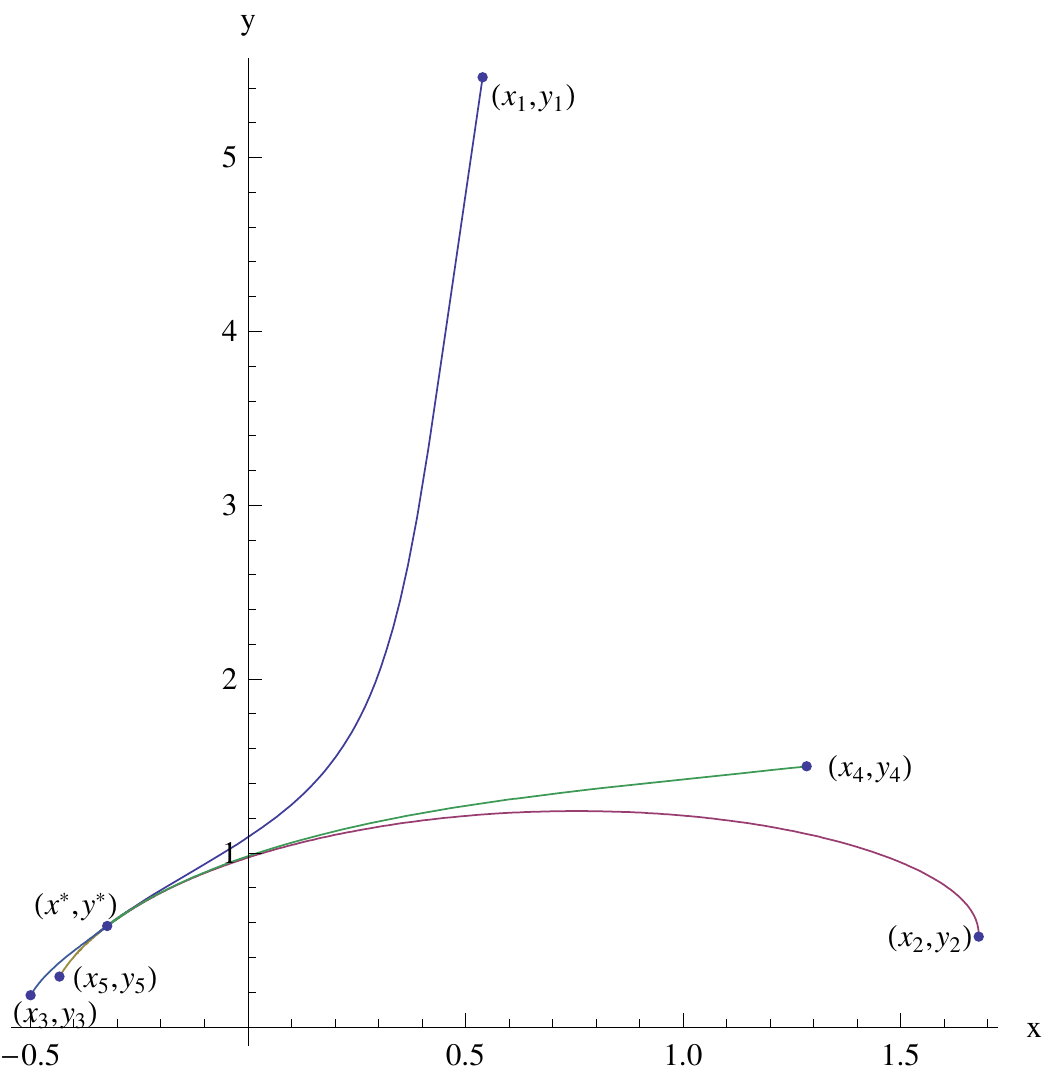}
  \caption{Non-BPS flow with charges $(p^0, p^1, q_1, q_0)= (5, 2, 7, 3)$ and attractor point $(x^*,y^*)=(-0.323385, 0.580375)$. The initial points of each flow are given by: $(x_1=0.539624, y_1= 5.461135), (x_2=1.67984, y_2=0.518725), (x_3=-0.432811,y_3=0.289493),(x_4=1.28447, y_4=1.49815), (x_5=-0.499491, y_5=0.181744)$.}
  \label{fig:NonBPSflow}
\end{figure}

A non-BPS attractor flow with generic charges and asymptotic moduli can be generated using the method detailed earlier. Figure \ref{fig:NonBPSflow} shows an example of non-BPS attractor flow with charges $(p^0, p^1, q_1, q_0)= (5, 2, 7, 3)$. Note that $J_4(5, 2, 7/3, 3)<0$, so this is indeed a non-BPS black hole.

Unlike the BPS attractor flows, all non-BPS flows starting from different asymptotic moduli reach the attractor point with the same tangent direction. The reason is, unlike the BPS case, the mass matrix of the black-hole potential $V_{BH}$ at a non-BPS critical point has two different eigenvalues. The common tangent direction for the non-BPS flows corresponds to the eigenvector associated with the smaller mass.

Now we discuss how to determine $k_{NB}$ and $g_{NB}$ for given D-brane charges and asymptotic moduli. Unlike the BPS case, there are only eight parameters in
$k_{NB}$ and $g_{NB}$: the two twistors $\{z,u\}$ and
$\{C^{\alpha a},G^{\alpha a}\}$ under the constraints $u=\frac{C^{22}}{C^{12}}=\frac{G^{22}}{G^{12}}$. On the other hand, there are still eight constraints in a given non-BPS attractor flow as in the BPS case. Therefore, while $k_{BPS}$ and $g_{BPS}$ can parameterize black holes with arbitrary $(p^I, q_I)$ and $(x_0, y_0)$ while leaving $\{z,u\}$ free, all the parameters in $k_{NB}$ and $g_{NB}$,
including $\{z,u\}$, will be fixed.

Another major difference from the BPS case is that
\begin{equation}[k_{NB},g_{NB}]=0\end{equation}
guaranteed by the form of $v^{NB}_a$ and $w^{NB}_a$. Thus the charge equation
(\ref{kfromQsingleg}) becomes simply
\begin{equation}\label{kfromQsingleNB}
\mathbf{Q}_{NB}=S_0 (k_{NB})S_0
\end{equation}

Unlike the BPS case, $g_{NB}$ does not enter the charge equations, thus cannot be used to eliminate the dependence on the twistor $z$. The three degrees of freedom in $g_{NB}$ are simply fixed by the asymptotic moduli $(x_0, y_0)$ and $u_0=1$, without invoking the zero NUT charge condition. The four D-brane charges equations in (\ref{kfromQsingleNB}) determine $C^{\alpha a}=C^{\alpha a}(Q,z)$,\footnote{See eq. (6.35) of \cite{Gaiotto:2007ag} for the full solution.} which then fixes $u$ via $u=\frac{C^{22}}{C^{12}}$. Finally, the zero NUT charge condition imposes a degree-six equation on twistor $z$:
\begin{equation}\label{zfromQNB}
p^0z^6+6p^1z^5-(3p^0+4q_1)z^4-4(3p^1-2q_0)z^3+(3p^0+4q_1)z^2+6p^1z-p^0=0
\end{equation}

Similar to the BPS case, the full non-BPS attractor flow can be generated
from the attractor moduli by replacing $C^{\alpha a}$ with the
harmonic function $H^{\alpha a}(\tau)=C^{\alpha a}\tau+G^{\alpha
a}$, while keeping $z$ fixed as in (\ref{Charmonic}):
\begin{equation}\label{CharmonicNB}
t_{NB}(\tau)= t_{NB}^{*}(C^{\alpha a}\rightarrow C^{\alpha a}\tau+G^{\alpha a},z)
\end{equation}
However, there are two important differences. First, the harmonic functions
$H^{\alpha a}$ have to satisfy the constraint:\footnote{This does not impose
any constraint on the allowed asymptotic moduli since there are
still three degrees of freedom in $G^{\alpha a}$ to account for
$(x_0, y_0, u_0)$. We will see later that its multi-centered counterpart helps impose a stringent constraint on the allowed D-brane charges in multi-centered
non-BPS solutions.}
\begin{equation}\label{NBHFconstraint}
\frac{H^{22}(\tau)}{H^{12}(\tau)}=u=\frac{C^{22}}{C^{12}}=\frac{G^{22}}{G^{12}}
\end{equation}

Second, unlike the BPS flow, a generic non-BPS flow cannot be given by
the ``naive" harmonic function procedure:
\begin{equation}\label{QharmonincProcedureNB}
t_{NB}(\tau)\neq t^{*}_{NB}(Q\rightarrow Q\tau+h)
\end{equation}
The reason is that the twistor $z$ in a non-BPS solution is
no longer free as in the BPS case, but is determined in terms of D-brane charges via (\ref{zfromQNB}). Thus replacing $Q$
with $Q\tau+h$, for generic $Q$ and $h$, would not leave $z$
invariant. That is, replacing $C^{\alpha a}$ in the
attractor moduli with harmonic functions $H^{\alpha a}(\tau)$ is
not equivalent to replacing the charges $Q$ with $H=Q\tau+h$ as in the BPS case
(\ref{QharmonincProcedure}).

It is interesting to find the subset of non-BPS single-centered flows that \textit{can} be constructed via the ``naive" harmonic function procedure. The $n_V=1$ system can be considered as the STU model with the three moduli $(S,T,U)$ identified. Since the STU model has an
$SL(2,\mathbb{Z})^3$ duality symmetry at the level of E.O.M., the $n_V=1$ system has an
$SL(2,\mathbb{Z})$ duality symmetry coming from identifying these
three $SL(2,\mathbb{Z})$'s, namely, $
\hat{\Gamma}=\left(\begin{array}{cc} a & b  \\
c & d  \end{array} \right)\otimes \left(\begin{array}{cc} a & b  \\
c & d  \end{array} \right)\otimes \left(\begin{array}{cc} a & b  \\
c & d  \end{array} \right)$ with $ad-bc=1$.
The modulus $t=x+i y$ transforms as $t\rightarrow \hat{\Gamma}t=\frac{a t+b}{c t+d}$,
and the transformation on the charges is given by
\cite{Behrndt:1996hu}.

Given an arbitrary charge $Q$, there exists a transformation
$\hat{\Gamma}_{Q}$ such that $Q=\hat{\Gamma}_{Q}Q_{40}$ for some
D4-D0 charge system $Q_{40}=(0,p^1,0,q_0)$. The solution of (\ref{zfromQNB}) with charge $Q=\hat{\Gamma}_{Q}Q_{40}$ has a root $z=\frac{a\pm \sqrt{a^2+c^2}}{c}$, independent of $Q_{40}$. Thus for arbitrary $h_{40}=(0,h^1,0,h_0)$, replacing $Q$ with $Q\tau +\hat{\Gamma}_{Q}h_{40}$ would leave the twistor $z$ invariant. We thus conclude
that the non-BPS single-centered attractor flows that can be
generated from their attractor moduli via the ``naive" harmonic function procedure are only
those with $(Q,h)$ being the image of a single transformation
$\hat{\Gamma}$ on a D4-D0 system $(Q_{40},h_{40})$:
\begin{equation}\label{QharmonincProcedureNB}
t_{NB}(\tau)=t^{*}_{NB}(\hat{\Gamma} Q_{40}\rightarrow
\hat{\Gamma}Q_{40}\tau+\hat{\Gamma}h_{40})
\end{equation}

\section{Multi-Centered Attractor Flows}

Similar to the single-centered attractor solutions,  the multi-centered ones are constructed by exponentiating harmonic matrix functions $K(\vec{x})$:
\begin{equation}
 S(\vec{x})=e^{K(\vec{x})}S_0
\end{equation}
Recall that for single-centered attractors, using the $\mathbf{H}$-action on $g$, $K(\tau)=k\tau+g$ can be adjusted to have the same properties as the flow generator $k$ as in (\ref{constraintonK}). For BPS multi-centered solutions, supersymmetry guarantees that the matrix function $K(\vec{x})$ also has the same properties as the generator $k$:
\begin{equation}\label{constraintonKmulti}
K^3(\vec{x})=0 \qquad \textrm{and} \qquad K^2(\vec{x})
\,\,\textrm{rank two}
\end{equation}
We will impose these constraints on all non-BPS multi-centered solutions as well, since presently we are more interested in the multi-centered solutions that are ``assembled" by individual single-centered attractors and thus have similar properties to their single-centered constituents. It is certainly interesting to see if there exist non-BPS multi-centered  solutions with $K(\vec{x})$ not sharing the constraints (\ref{constraintonKmulti}) satisfied by the flow generator $k_{NB}$.

The harmonic matrix function $K(\vec{x})$ satisfying all the above constraints is
solved to be:
\begin{equation}
K(\vec{x})=\sum_i\frac{k_i}{|\vec{x}-\vec{x}_i|}+g
\end{equation}
where
\begin{equation}
k_i=\sum_{a=1,2} [v_a (w_a)_i^T+ (w_a)_i v_a^T]S_0  \qquad
\textrm{and} \qquad g=\sum_{a=1,2}[v_a m_a^T+ m_a v_a^T] S_0
\end{equation}
with $v_a$ being the same two constant null vectors in single-centered $k$, and
the 7-vectors $(w_a)_i$ contain the information of the D-brane charges of center-$i$, and the two 7-vectors $m_a$'s contain that of asymptotic moduli. Both $(w_a)_i$ and $m_a$ are orthogonal to $v_b$. Since $v_a$ only depends on the twistor $\{z,u\}$, and $w_a$
are linear in $C^A$ or $C^{\alpha a}$, the above generating
procedure is equivalent to replacing $C^{A}$ and $C^{\alpha a}$ with
the multi-centered harmonic functions $H^{A}(\vec{x})$ and $H^{\alpha a}(\vec{x})$ while keeping the twistor $\{z,u\}$ fixed.

\subsection{Multi-Centered BPS Attractors}

Using $\mathbf{Q}_i$ to denote the charge matrix of center-$i$, we have $5N$ coupled
equations from $\mathbf{Q}_i=\frac{1}{4\pi}\int_i \nabla \cdot J$:
\begin{equation}\label{kfromQmulti}
\mathbf{Q}_{BPS,i}=S_0 (k_{BPS,i}+\frac{1}{2}[k_{BPS,i},
g_{BPS}]+\frac{1}{2}\sum_j\frac{[k_{BPS,i},k_{BPS,j}]}{|\vec{x}_i-\vec{x}_j|})S_0
\end{equation}

We now show in detail how to determine $k_{BPS,i}$ and $g_{BPS}$ for given charges and asymptotic moduli using equation (\ref{kfromQmulti}). There are $4(N+1)+1$ parameters in $k_{BPS,i}$ and $g_{BPS}$: $\{C^A_i, G^A,z\}$,
since the twistor $u$ is arbitrary. Different from the single-centered BPS case, there are also $3N-3$ degrees of freedom from
the positions of centers on L.H.S. of (\ref{kfromQmulti}). On the other hand, there are $5N+3$ constraints in a given BPS multi-centered attractor: $4N$
D-brane charges $(p^I_i,q_{I,i})$, $N$ vanishing NUT charges, the asymptotic moduli
$(x_0, y_0)$ and $u_0=1$.
We will use these $5N+3$ constraints to fix the $4(N+1)$ parameters $\{C^A_i, G^A\}$ in $k_{BPS,i}$ and $g_{BPS}$, and impose $N-1$ constraints on the distances between the $N$ centers, while leaving the twistor $z$ free.

First, integrating $\nabla \cdot J$ over the sphere at the infinity gives the sum of
the above $N$ matrix equations: $\mathbf{Q}^{tot}_{BPS}=S_0(k^{tot}_{BPS}+\frac{1}{2}[k^{tot}_{BPS}, g_{BPS}])S_0$,
which is the same as the charge equation for a single-center attractor with
charge $Q^{tot}_{BPS}$. This determines $g$ to be $g=g_h+\Lambda$, same
as the single-centered case as in (\ref{gghlambda}), using the three asymptotic moduli $(x_0, y_0,u_0)$ and the constraint of zero total NUT charge. The $h$'s are fixed by the asymptotic moduli and the integrability condition $<Q^{tot}_{BPS},h>=0$.

It is easy to see that the solutions of $C^A_i$ are simply given by the single-centered solutions $C^A=C^A(Q,z)$ with $Q$ replaced by
$Q_i$. Thus the flow generator of each center $k_{BPS,i}$ (given by $k_{BPS,i}=a_{\alpha A}z^{\alpha}C^{A}_i$) satisfies
\begin{equation}\label{kfromQmulti-i}
\mathbf{Q}_{BPS,i}=S_0 (k_{BPS,i}+\frac{1}{2}[k_{BPS,i}, \Lambda])S_0
\end{equation}
which is the multi-centered generalization of the single-centered condition (\ref{kfromQsinglelambda}).

Using the solutions of $k_{BPS,i}$ and $g_{BPS}$, the charge equations (\ref{kfromQmulti}) become
\begin{equation}
\mathbf{Q}_{BPS,i}=S_0 \left(k_{BPS,i}+\frac{1}{2}[k_{BPS,i},
\Lambda]+(<Q_{BPS,i}, h>+\sum_j\frac{<Q_{BPS,i}, Q_{BPS,j}>}{|\vec{x}_i-\vec{x}_j|})\Theta\right)S_0
\end{equation}
from which we subtract (\ref{kfromQmulti-i}) to produce the
integrability condition
\begin{equation}\label{integrabilitymulti}
<Q_{BPS,i},h>+\sum_j\frac{<Q_{BPS,i},Q_{BPS,j}>}{|\vec{x}_i-\vec{x}_j|}=0
\end{equation}

The sum of the $N$ equations in the integrability condition
(\ref{integrabilitymulti}) reproduces the constraint on $h$:
$<Q^{tot}_{BPS},h>=0$. Thus the remaining $N-1$ equations impose $N-1$
constraints on the relative positions between the $N$ centers. The angular momentum $\vec{J}$, defined via $\omega_i = 2\epsilon_{ijk} J^j
\frac{x^k}{r^3}$ as  $r\rightarrow \infty$, is non-zero:
\begin{equation}
\vec{J}=\frac{1}{2}\sum_{i<j}\frac{\vec{x}_i-\vec{x}_j}{|\vec{x}_i-\vec{x}_j|}\langle
Q_{BPS,i},Q_{BPS,j}\rangle
\end{equation}
Thus we have shown that our multi-centered BPS attractor solutions
reproduce those found in \cite{Bates:2003vx}.
Same arguments as in the single-centered BPS case shows that multi-centered
BPS attractors can be generated by replacing the charges in the
attractor moduli with corresponding multi-centered harmonic functions:
\begin{equation}
t_{BPS}(\vec{x})=t^{*}_{BPS}(Q_{BPS}\rightarrow
\sum_i\frac{Q_{BPS,i}}{|\vec{x}-\vec{x}_i|}+h)
\end{equation}

\subsection{Multi-Centered Non-BPS Attractors}

A multi-centered non-BPS attractor has  $3(N+1)+2$
parameters inside its non-BPS generators $\{k_{NB,i},g_{NB}\}$: $\{C^{\alpha a}_i,
G^{\alpha a}\}$ under the constraint (\ref{NBCGconstraintmulti}) plus two twistors $\{z,u\}$. Given $\{k_{NB,i},g_{NB}\}$ in terms of $\{C^{\alpha a}_i,G^{\alpha a},z,u\}$, the
non-BPS multi-centered solution is the same as the single-centered
one with $H^{\alpha a}(\tau)$ replaced by
multi-centered harmonic functions $H^{\alpha a}(\vec{x})=\sum_i
\frac{C^{\alpha a}_i}{|\vec{x}-\vec{x}_i|}+G^{\alpha a}$
satisfying the constraint
\begin{equation}\label{NBCGconstraintmulti}
u=\frac{H^{22}_i(\vec{x})}{H^{12}_i(\vec{x})}=\frac{C^{22}_i}{C^{12}_i}=\frac{G^{22}}{G^{12}}
\end{equation}
However, the process of determining $k_{NB,i}$ and $g_{NB}$ in terms of charges and asymptotic moduli for a non-BPS multi-centered attractor is very different from its BPS counterpart.

The reason is that the charge equations for a non-BPS multi-centered solution
simplifies a great deal since
\begin{equation}
[k_{NB,i},k_{NB,j}]=0 \qquad \textrm{and} \qquad [k_{NB,i},g_{NB}]=0
\end{equation}
guaranteed by the forms of $(w^{NB}_a)_i$ and
$m^{NB}_{a}$. Therefore, the $5N$ equations (\ref{kfromQmulti})
decouple into $N$ sets of $5$ coupled equations:
\begin{equation}\label{kfromQmultiNB}
\mathbf{Q}_{NB,i}=S_0 (k_{NB,i})S_0
\end{equation}
As in the single-centered non-BPS case, $g_{NB}$ does not enter the charge equations (\ref{kfromQmultiNB}), and its three degrees of freedom can be completely fixed by the given asymptotic moduli $(x_0,y_0)$ and $u_0=1$ without using the zero NUT charge condition. More importantly, unlike BPS multi-centered solutions, the
positions of centers $\vec{x}_i$ do not appear in the charge equations (\ref{kfromQmultiNB}), thus receive no constraint: all centers are free.
Finally, since we are using the remaining $3N+2$ parameters $\{C^{\alpha a},z,u\}$ to parameterize a $N$-centered attractor solution under $5N$ constraints coming from charge equations (\ref{kfromQmultiNB}), there need to be $2N-2$ constraints imposed on the D-brane charges.

As in the BPS multi-centered attractors, solutions of $C^{\alpha a}_i$ are given by the single-centered non-BPS solutions $C^{\alpha a}=C^{\alpha a}(Q,z)$ with $Q$ replaced by
$Q_i$. The solutions of twistors $z$ and $u$ are the same as the single-centered ones with charges $Q_{NB}$ replaces by $Q^{tot}_{NB}$. Among the aforementioned $2N-2$ constraints, $N-1$ come from demanding that all centers have the same twistor $z$, which follows from the zero NUT charge condition at each center, and the other $N-1$ come from demanding that they  have the same twistor $u$ as in (\ref{NBCGconstraintmulti}). Solving these $2N-2$ constraints shows that all the charges $\{Q_{NB,i}\}$ are the
image of a single duality transformation $\hat{\Gamma}$ on a
multi-centered D4-D0 system $\{Q_{NB,40,i}\}$:
\begin{equation}
Q_{NB,i}=\hat{\Gamma}Q_{NB,40, i}
\end{equation}
The charges at different centers are all mutually local
\begin{equation}
\langle Q_{NB,i}, Q_{NB,j} \rangle =0
\end{equation}

Like non-BPS single-centered attractors, the generic non-BPS multi-centered attractors cannot be generated via the ``naive" harmonic function procedure, except for those with $\{ Q_{NB,i}, h \}$ being the image of a single
$\hat{\Gamma}$ on a pure D4-D0
system $\{Q_{NB,40,i}, h_{40}\}$:
\begin{equation}\label{QharmonincProcedureNBmulti}
t_{NB}(\vec{x})=t^{*}_{NB}(\hat{\Gamma} Q_{NB,40}\rightarrow
\sum_i\frac{\hat{\Gamma}Q_{NB,40,i}}{|\vec{x}-\vec{x}_i|}+\hat{\Gamma}h_{40})
\end{equation}

In summary, the non-BPS multi-centered attractors are drastically different
from their BPS counterparts: there is no constraint imposed on the positions of the centers, but instead on the allowed charges $Q_{NB,i}$: they have to be mutually local.
The result is that the centers can move freely, and there is no intrinsic angular momentum in the system.

\section{Conclusion and Discussion}

In this talk, we summarized the construction of generic single-centered and
multi-centered extremal black hole solutions in theories whose 3D moduli spaces are symmetric coset spaces. In this construction, all attractors, both BPS and non-BPS, single-centered as well as multi-centered, are treated on an equal footing.
The single-centered black hole attractors correspond to those null geodesics in $\mathcal{M}_{3D}$ that are generated by exponentiating appropriate nilpotent elements in the coset algebra. The multi-centered black hole attractors are given by $3D$ solutions that live in certain null totally geodesic sub-manifolds of $\mathcal{M}_{3D}$. The construction of multi-centered attractors, even that of non-BPS ones, is merely a straightforward generalization of the single-centered construction.

We presented a detailed computation in the theory of 4D $\mathcal{N}=2$ supergravity coupled to one vector-multiplet, whose 3D moduli space is the symmetric coset space $G_{2(2)}/SL(2,\mathbb{R})^2$. The attractor flow generators are third-degree nilpotent elements in the coset algebra. We explicitly constructed generic attractor solutions, both single-centered and multi-centered, and showed that while the BPS attractors can be generated from the attractor moduli via the ``naive" harmonic function procedure, the generic non-BPS attractors cannot be generated this way.

In the $n_V=1$ model, besides the BPS generator, there is only one extra third-degree nilpotent orbit to serve as non-BPS flow generators. Hence there is only one type of non-BPS single-centered attractor. In models with bigger symmetric moduli spaces, there should be more than one type of non-BPS generator. These would give rise to different types of non-BPS attractor flows, which might have different stability properties.

All multi-centered non-BPS attractors constructed in this work follow from the ansatz in which 3D gravity is assumed to decouple from the moduli. The multi-centered non-BPS black holes are found to be very different from their BPS counterparts: the charges of all centers are constrained to be mutually local, while the positions of centers are completely free. Thus the non-BPS multi-centered attractor is not a ``bound state" and carries no intrinsic angular momentum.

We would like to construct true multi-centered non-BPS ``bound states", i.e., solutions with constraints on the positions of centers but not on the charges. There are two possible ways to achieve this. First, one could adopt a more general ansatz in which 3D gravity is coupled to the moduli. For axisymmetric configurations, the inverse scattering method could be used to perform an exact analysis. One could also search in models with bigger moduli spaces. It is very likely that in bigger moduli spaces, there exist true multi-centered non-BPS ``bound states" even within the ansatz with 3D gravity decoupled from moduli. We are also interested in the possibility of generating multi-centered non-BPS solutions with each center having different types of non-BPS generators $k_{NB}$.

Finally, with the hope of studying non-BPS extremal black holes in 4D ${\cal N}=2$ supergravity coupled to $n_V$ vector-multiplets with more generic pre-potential, we would like to generalize our method to non-symmetric homogeneous spaces, and even to generic moduli spaces eventually.

\section*{Acknowledgements}

The author would like to thank D. Gaiotto and M. Padi for the collaboration on the project this talk was based on. We are grateful to A. Neitzke and J. Seo for helpful discussions. It is also a pleasure to thank the warm hospitality of the School of Attractor Mechanism (2007). The work is supported by DOE grant DE-FG02-91ER40654.

%

\begin{thebibliography}{10}






\bibitem{Ferrara:1995ih}
  S.~Ferrara, R.~Kallosh and A.~Strominger,
  Phys.\ Rev.\  D {\bf 52}, 5412 (1995)
  [arXiv:hep-th/9508072].

\bibitem{Sen:2005wa}
  A.~Sen,
  JHEP {\bf 0509}, 038 (2005)
  [arXiv:hep-th/0506177].





\bibitem{Goldstein:2005hq}
  K.~Goldstein, N.~Iizuka, R.~P.~Jena and S.~P.~Trivedi,
  Phys.\ Rev.\  D {\bf 72}, 124021 (2005)
  [arXiv:hep-th/0507096].


\bibitem{Sen:2005iz}
  A.~Sen,
  JHEP {\bf 0603}, 008 (2006)
  [arXiv:hep-th/0508042].

\bibitem{Tripathy:2005qp}
  P.~K.~Tripathy and S.~P.~Trivedi,
  JHEP {\bf 0603}, 022 (2006)
  [arXiv:hep-th/0511117].






\bibitem{Alishahiha:2006ke}
  M.~Alishahiha and H.~Ebrahim,
  JHEP {\bf 0603}, 003 (2006)
  [arXiv:hep-th/0601016].

\bibitem{Kallosh:2006bt}
  R.~Kallosh, N.~Sivanandam and M.~Soroush,
  JHEP {\bf 0603}, 060 (2006)
  [arXiv:hep-th/0602005].


\bibitem{Chandrasekhar:2006kx}
  B.~Chandrasekhar, S.~Parvizi, A.~Tavanfar and H.~Yavartanoo,
  JHEP {\bf 0608}, 004 (2006)
  [arXiv:hep-th/0602022].

\bibitem{Sahoo:2006rp}
  B.~Sahoo and A.~Sen,
  JHEP {\bf 0609}, 029 (2006)
  [arXiv:hep-th/0603149].


\bibitem{Astefanesei:2006dd}
  D.~Astefanesei, K.~Goldstein, R.~P.~Jena, A.~Sen and S.~P.~Trivedi,
  JHEP {\bf 0610}, 058 (2006)
  [arXiv:hep-th/0606244].

\bibitem{Kallosh:2006ib}
  R.~Kallosh, N.~Sivanandam and M.~Soroush,
  Phys.\ Rev.\  D {\bf 74}, 065008 (2006)
  [arXiv:hep-th/0606263].



\bibitem{Sahoo:2006pm}
  B.~Sahoo and A.~Sen,
  JHEP {\bf 0701}, 010 (2007)
  [arXiv:hep-th/0608182].

\bibitem{Andrianopoli:2006ub}
  L.~Andrianopoli, R.~D'Auria, S.~Ferrara and M.~Trigiante,
  arXiv:hep-th/0611345.

\bibitem{D'Auria:2007ev}
  R.~D'Auria, S.~Ferrara and M.~Trigiante,
  JHEP {\bf 0703}, 097 (2007)
  [arXiv:hep-th/0701090].




\bibitem{Nampuri:2007gv}
  S.~Nampuri, P.~K.~Tripathy and S.~P.~Trivedi,
  arXiv:0705.4554 [hep-th].




\bibitem{Dabholkar:2006tb}
  A.~Dabholkar, A.~Sen and S.~P.~Trivedi,
  JHEP {\bf 0701}, 096 (2007)
  [arXiv:hep-th/0611143].

\bibitem{Saraikin:2007jc}
  K.~Saraikin and C.~Vafa,
  arXiv:hep-th/0703214.



\bibitem{Ferrara:1997tw}
  S.~Ferrara, G.~W.~Gibbons and R.~Kallosh,
  Nucl.\ Phys.\  B {\bf 500}, 75 (1997)
  [arXiv:hep-th/9702103].



\bibitem{Bates:2003vx}
  B.~Bates and F.~Denef,
  arXiv:hep-th/0304094.

\bibitem{Denef:2007vg}
  F.~Denef and G.~W.~Moore,
  arXiv:hep-th/0702146.

\bibitem{Gaiotto:2007ag}
  D.~Gaiotto, W.~W.~Li and M.~Padi,
  arXiv:0710.1638 [hep-th].


\bibitem{Ceresole:2007wx}
  A.~Ceresole and G.~Dall'Agata,
  JHEP {\bf 0703}, 110 (2007)
  [arXiv:hep-th/0702088].


\bibitem{Lopes Cardoso:2007ky}
  G.~Lopes Cardoso, A.~Ceresole, G.~Dall'Agata, J.~M.~Oberreuter and J.~Perz,
  JHEP {\bf 0710}, 063 (2007)
  [arXiv:0706.3373 [hep-th]].



  
  
\bibitem{Breitenlohner:1987dg}
  P.~Breitenlohner, D.~Maison and G.~W.~Gibbons,
  Commun.\ Math.\ Phys.\  {\bf 120}, 295 (1988).
  
\bibitem{Cecotti:1988qn}
  S.~Cecotti, S.~Ferrara and L.~Girardello,
  Int.\ J.\ Mod.\ Phys.\  A {\bf 4}, 2475 (1989).
  
\bibitem{Ferrara:1989ik}
  S.~Ferrara and S.~Sabharwal,
  Nucl.\ Phys.\  B {\bf 332}, 317 (1990).


\bibitem{de Wit:1992wf}
  B.~de Wit, F.~Vanderseypen and A.~Van Proeyen,
  Nucl.\ Phys.\  B {\bf 400}, 463 (1993)
  [arXiv:hep-th/9210068].

\bibitem{Ceresole:1995ca}
  A.~Ceresole, R.~D'Auria and S.~Ferrara,
  Nucl.\ Phys.\ Proc.\ Suppl.\  {\bf 46}, 67 (1996)
  [arXiv:hep-th/9509160].


\bibitem{Gunaydin:2005mx}
  M.~Gunaydin, A.~Neitzke, B.~Pioline and A.~Waldron,
  Phys.\ Rev.\  D {\bf 73}, 084019 (2006)
  [arXiv:hep-th/0512296].



\bibitem{Pioline:2006ni}
  B.~Pioline,
  Class.\ Quant.\ Grav.\  {\bf 23}, S981 (2006)
  [arXiv:hep-th/0607227].




\bibitem{Bouchareb:2007ax}
  A.~Bouchareb, G.~Clement, C.~M.~Chen, D.~V.~Gal'tsov, N.~G.~Scherbluk and T.~Wolf,
  Phys.\ Rev.\  D {\bf 76}, 104032 (2007)
  [arXiv:0708.2361 [hep-th]].

\bibitem{Clement:2007qy}
  G.~Clement,
  arXiv:0710.1192 [gr-qc].

\bibitem{Gunaydin:2007qq}
  M.~Gunaydin, A.~Neitzke, O.~Pavlyk and B.~Pioline,
  arXiv:0707.1669 [hep-th].





\bibitem{Behrndt:1996hu}
  K.~Behrndt, R.~Kallosh, J.~Rahmfeld, M.~Shmakova and W.~K.~Wong,
  Phys.\ Rev.\  D {\bf 54}, 6293 (1996)
  [arXiv:hep-th/9608059].





\bibitem{Collingwood:1993}
  D.~H. ~Collingwood and W.~M. ~McGovern,
  Nilpotent orbits in semisimple Lie algebras Van Nostrand Reinhold,
  New York U.S.A. (1993).
  
  





\end{thebibliography}
%



\printindex
\end{document}